\RequirePackage{fix-cm}
\documentclass[twocolumn]{svjour3}          
\smartqed  
\usepackage{array}
\usepackage{amsmath}
\usepackage{amsbsy}
\usepackage{amssymb}
\usepackage{algorithm}
\usepackage{algpseudocode}
\usepackage{float}
\usepackage{listings}
\usepackage{url}
\usepackage{hyperref}
\usepackage{graphicx}
\usepackage{enumitem}
\usepackage{tikz}
\usepackage{subcaption}
\usepackage{mathrsfs}
\usepackage{breqn}
\usepackage[numbers]{natbib}

\usepackage{tikz-cd}

\def\middlebreak {\nulldelimiterspace0pt
\right.\allowbreak\mskip 0mu plus .5mu \nulldelimiterspace0pt\left.}%

\newcommand{\FILE}[1]{\textrm{file}_{#1}}
\newcommand{\PNAME}[1]{\textrm{pname}_{#1}}
\newcommand{\PVAL}[1]{\textrm{pval}_{#1}}
\newcommand{\SPAIR}[1]{\textrm{sess}_{#1}}

\newcommand{\HTTPHEADER}[1]{\textrm{header}_{#1}}

\newcommand{\READ}[1]{\mathtt{read(#1)}}
\newcommand{\DEEPREAD}[1]{\mathtt{deepread(#1)}}
\newcommand{\SEARCH}[1]{\mathtt{search(#1)}}

\newcommand{\GET}[1]{\mathtt{get(#1)}}
\newcommand{\POST}[1]{\mathtt{post(#1)}}

\newcommand{\READi}{\mathtt{read}(\textrm{file}_{i})}
\newcommand{\SEARCHi}{\mathtt{search}(\textrm{file}_{i})}

\newcommand{\DEEPREADii}{\mathtt{deepread}(\textrm{file}_{i}}

\newcommand{\READiii}{\mathtt{read}(\textrm{file}_{i},\middlebreak \textrm{pname}_{j},\middlebreak \textrm{pval}_{k} )}
\newcommand{\SEARCHiii}{\mathtt{search}(\textrm{file}_{i},\middlebreak \textrm{pname}_{j},\middlebreak \textrm{pval}_{k} )}
\newcommand{\DEEPREADiii}{\mathtt{deepread}\middlebreak(\textrm{file}_{i},\middlebreak \textrm{pname}_{j},\middlebreak \textrm{pval}_{k} )}

\newcommand{\GETiv}{\mathtt{get}(\textrm{file}_{i}, \middlebreak [\textrm{pname}], \middlebreak [\textrm{pval}] )}
\newcommand{\POSTiv}{\mathtt{post}(\textrm{file}_{i}, \middlebreak [\textrm{pname}, \middlebreak [\textrm{pval}] )}

\newcommand{\GETv}{\mathtt{get}(\textrm{file}_{i}, \middlebreak [\textrm{pname}], \middlebreak [\textrm{pval}], \middlebreak [\textrm{sess}], \middlebreak \textrm{header} )}
\newcommand{\POSTv}{\mathtt{post}(\textrm{file}_{i}, \middlebreak [\textrm{pname}], \middlebreak [\textrm{pval}], \middlebreak [\textrm{sess}], \middlebreak \textrm{header} )}

\graphicspath{{./img/}}

\begin{document}

\title{The Agent Web Model 
}
\subtitle{Modelling web hacking for reinforcement learning}


\author{L{\'a}szl{\'o} Erd{\H o}di         \and
        Fabio Massimo Zennaro 
}


\institute{L. Erd{\H o}di \at
              Department of Informatics,\\
              University of Oslo, \\
              0316 Oslo, Norway \\
              \email{laszloe@ifi.uio.no}           
           \and
           F.M. Zennaro \at
              Department of Informatics,\\
              University of Oslo, \\
              0316 Oslo, Norway \\
              \email{fabiomz@ifi.uio.no}
}

\date{Received: date / Accepted: date}

\maketitle

\begin{abstract}
Website hacking is a frequent attack type used by malicious actors to obtain confidential information, modify the integrity of web pages or make websites unavailable. The tools used by attackers are becoming more and more automated and sophisticated, and malicious machine learning agents seems to be the next development in this line. In order to provide ethical hackers with similar tools, and to understand the impact and the limitations of artificial agents, we present in this paper a model that formalizes web hacking tasks for reinforcement learning agents. Our model, named \textit{Agent Web Model}, considers web hacking as a capture-the-flag style challenge, and it defines reinforcement learning problems at seven different levels of abstraction. We discuss the complexity of these problems in terms of actions and states an agent has to deal with, and we show that such a model allows to represent most of the relevant web vulnerabilities. Aware that the driver of advances in reinforcement learning is the availability of standardized challenges, we provide an implementation for the first three abstraction layers, in the hope that the community would consider these challenges in order to develop intelligent web hacking agents.

\keywords{Agent Web Model \and Penetration Testing \and Capture the Flag \and Reinforcement Learning}
\end{abstract}

\section{Introduction}\label{sec:introduction}
As the complexity of computer systems and networks significantly increased during the last decades, the number of vulnerabilities inside a system similarly raised. Different types of attackers may try to exploit these varying vulnerabilities for their own benefits. Websites are especially of interest to malicious actors, so attacks against websites nowadays are an everyday event. 
In order to protect vulnerable systems, one of the best approaches is to emulate real attacks using the same methodology that hackers would use. This practice, named \emph{white hat hacking}, has become a crucial part of critical information technology projects.
When taking part into a white hat hacking project aimed at testing the security of a target website, ethical hackers attack the system and report all their findings to the system owner or administrator so that the vulnerabilities can be patched. 
Ethical hacking is normally a human job, since the attacker needs a high level of expertise in penetration testing, which involves typically human capabilities (such as experience, reasoning, or intuition) that are hard to codify.

Although full automation of penetration testing is very challenging, hackers rely on a range of automatic tools \cite{fonseca2007testing}   \cite{NessusAudting} \cite{AntunesDesigning} to help them dealing with the number and the variety of possible vulnerabilities. 
In the case of web testing, there are many web security scanners that can help the work of a human tester. These tools can use predefined requests to check the existence of a vulnerability, and quickly generate security reports; however they have limited capability to carry out complex evaluations, and their findings must normally be reviewed by a human supervisor. Indexes of quality, such as the number of false positives and false negatives, highlight the limited coverage of these tools. New vulnerability detection scripts and general updates may be deployed to improve the performance of web vulnerability scanners, but these are usually one-time solutions; automatic improvements, relying, for instance, on learning from previous cases, are lacking. 
Furthermore, many web scanners are designed only to detect vulnerabilities, but not to exploit them. Specific tools can be used to exploit \cite{Sqlmap}, with a moderate chance of success, targeted vulnerabilities, and thus further the understanding of the overall security of the system under study. 

Machine learning (ML) techniques aimed at solving problems through learning and inference are now being adopted in many fields, including security \cite{StasinopoulosAutomatic}. Following their success in challenging tasks like image recognition \cite{krizhevsky2012imagenet} or natural language processing \cite{vaswani2017attention}, supervised deep neural network models have been adopted to tackle security-related problems in a static context, such as program vulnerability detection \cite{russell2018automated} or malicious domain name detection \cite{lison2017automatic}. However deep neural networks designed to solve static problems exploiting large data sets of examples do not conform to the more complex and dynamic problem of penetration testing. 
A sub-field of ML that may offer a more relevant paradigm to tackle problems such as web testing, is reinforcement learning. Indeed, reinforcement learning methods allow an agent to learn by itself in a dynamic and complex environment by trial-and-error and inference. Success on challenging games like Go \cite{silver2017mastering} or Starcraft II \cite{vinyals2019grandmaster} suggests that these algorithm may find soon use in the world of penetration testing.
Recently, some applications of ML and reinforcement learning in the context of offensive security were developed; on the side of white hat hackers, DARPA organized in 2016 the Cyber Grand Challenge for automated penetration testing \cite{CyberGrand}; on the side of black hat hackers malicious bots are being provided with more learning functionalities.

Given the impact that artificial agents will have in the landscape of security, this paper aims at promoting research in this direction by proposing a modelling of penetration testing problems that may be used to train reinforcement learning agents. Our modelling effort follows two directions: we first examine the \emph{formalization} of web hacking problems using standard models, and we then discuss \emph{abstractions} of concrete instances of web hacking problems within our model. We call our generic model the \emph{Agent Web Model}. 
Aware that a strong and effective driver for the development of new and successful reinforcement learning agents is the availability of standardized challenges and benchmark, we use our formalization to implement a series of challenges at different level of abstractions and with increasing complexity. We make these challenges available following the standards of the field. Our hope is that these challenges will promote and advance research in the development of automatic red bots that may help in the tasks of penetration testing.

This paper is organized as follows. Section \ref{sec:Background} presents the main concepts related to web hacking and reinforcement learning. Section \ref{sec:Formalization} discusses how the generic problem of web hacking may be reduced, through a set of formalization steps, to a reinforcement learning problem. Section \ref{sec:AgentWebModel} describes our own model for web hacking problems and describes instances of problems at different level of abstraction. Section \ref{sec:ModellingWeb} explains how real-world hacking problems may be mapped onto the Web Agent Model. Section \ref{sec:Implementation} provides some details on the implementation of challenges based on our formalization. Finally, Section \ref{sec:Ethical} discusses some ethical considerations about this work, and Section \ref{sec:Conclusions} draws conclusions and illustrates possible directions for future work.

\section{Background \label{sec:Background}}

\subsection{Web hacking}
The most famous and popular Internet service, the World Wide Web (WWW), has been running for many years \cite{BernersOriginalWeb}. Since its invention in 1989 it had undergone many developments, and nowadays it is one of the most complex services on the Internet. The HTTP protocol \cite{HTTP} used by these web services has been created for the communication within a client-server model. The web client, typically a web browser, sends a HTTP request to a webserver; the webserver, in turn, answers with a HTTP response. 
A HTTP messages consist of three main parts: the Uniform Resource Locator (URL), the HTTP header, and the HTTP body. The URL references the requested object. The header contains information on the state of the communication. The request header sent by a client specifies the web method (i.e., what to do with the object), client-related information (e.g., the type of the web browser), and cookie values referring to previous states of the communication. The answer header sent by a server contains the answer code to the request (e.g., file not found) and information related to the state of the communication (e.g., new cookie values with session variables).
The body part of the HTTP message contains the payload of the communication. The request body may contain POST parameters sent by the client. The answer body usually contains the longest part of the message, that is, the web page content in Hypertext Markup Language (HTTP) format.

Web communication is well defined by the HTTP standard. In time, due to the high number of components participating in the web communication, the web protocol has become increasingly complex, opening room to different vulnerabilities \cite{OwaspWeb}. 
On the client side, a minimal web client can be easily realized by instantiating a TCP connection and by sending HTTP requests via command line. However, to enjoy the rich functionalities provided by HTML, including the last standard HTML5, web browsers are normally used. A web browser is an application providing a graphical interface that shows a HTML page with all its components. A HTML page may also contain code in the form of client-side scripts (such as javascript or scripts in other embedded objects), which can be executed locally by a web browser. Any unintended or malicious client-side script can have serious consequences during the web communication. If an attacker can sniff web traffic without encryption, or if the message can be decrypted, then attackers can set up man-in-the-middle exploitations. 
On the server side, the webserver runs on a physical or virtual computer, which may expose non-HTTP-related vulnerabilities at the level of the operating systems, at the level of the applications running the webserver (e.g., Apache, Ngin-x, IIS) or by exposing other vulnerable services on the web. With respect to the HTTP protocol, one of the most significant, and potentially vulnerable, parts of website access is the server-side scripting engine. Server-side scripting makes possible for the server to accept input sent by the client in order to customize its web answer. Based on the input, a server-side script can create connections to other resources such as local files or database records. Many components, such as Content Management Systems (CMS), provide ready modules for different functionalities using some server-side scripts. Having a vulnerability in a CMS can expose millions of website that run the same vulnerable module.
 
Using the web protocol can thus expose several weak points that can be targeted by malicious actors. The type of the attacks can vary, but they can be categorized according to the information security triplet. Several attacks aim to break the \emph{confidentiality} by accessing sensitive or confidential information; in such attacks, the attacker may be able to find hidden objects such as files or database data, or she may manage to escalate her privileges in order to access protected data. In other cases, object \emph{integrity} is targeted, either to cause damage and annoyance or as a preparatory step before carrying out further action; for instance, an  attacker may upload a command script to the website (changing the integrity of the site) and use it to further her attack with more options. The third type of attack addresses the \emph{availability} of the service; overloading a web service with many request can cause a denial of service (DOS). 

\subsection{Capture the Flag} \label{ssec:CTF} 
A Capture The Flag challenge (CTF) is a competition designed to offer to ethical hackers a platform to learn about penetration testing and train their skills \cite{mcdaniel2016capture}. CTFs are organized as a set of well-formalized and well-defined hacking challenges. Each challenge has one exploitable vulnerability (or, sometimes, a chain of vulnerabilities) and an unambiguous victory condition in the form of a \emph{flag}, that is, a token that proves whether the challenge was solved or not. Usually, a CTF requires purely logical and technical skills, and they exclude reliance on side channels such as social engineering; moreover, challenges are normally designed to make the use of brute-forcing or automatic tools unfeasible.

The standard setup of a CTF is the so-called \emph{Jeopardy mode}, in which all players target a single static system. More realistic setups may include the deployment of non-static services with evolving vulnerabilities, or the partition of players in teams, usually a \emph{red team}, tasked with retrieving flags from the target system, and \emph{blue team}, responsible for preventing the attacker from obtaining the flags.

In the case of web challenges, a standard CTF consists of a website hosting objects with different vulnerabilities, and containing flags in the form of special strings. Participants are required simply to collect the flag, and no further exploitative actions are required (such as, setting up a command and control system).
Jeopardy-style web CTFs constitute collections of rigorous challenges: the environment in which to operate is well-defined, actions can take place only in the digital domain, and objectives and victory conditions are clearly stated. All these properties make CTFs interesting case-studies to develop artificial agents for penetration testing.

\subsection{Reinforcement Learning}
Reinforcement learning (RL) is a sub-field of machine learning focused on the training of agents in a given environment \cite{sutton2018reinforcement}. 
Within such an environment, agents are given the possibility to choose actions from a finite set of available actions; upon undertaking an action, they can observe the consequences of their actions, both in terms of the effect on the environment, and in terms of a reward signal that specify how good or desirable is the outcome of that action. The aim of RL is to define algorithms that would allow an agent to develop an action policy leading to as high a reward as possible in time. 

The RL problem may be particularly challenging, as the space of actions for the agent may be large, the environment may be stochastic and non-stationary, and the reward signal may be sparse. However, despite these difficulties, RL has been proved successful in tackling a wide range of problems, such as mastering games \cite{mnih2015human,silver2017mastering} or driving autonomous vehicles \cite{shalev2016safe}. The ability to learn in complex learning environment, such as Starcraft II \cite{vinyals2019grandmaster}, mirrors the sort of learning that a web hacking agent is expected to perform. RL algorithms may then offer a way to train artificial agent able to carry out meaningful penetration testing.

\subsection{Related Work}
Interest in training artificial red bots able to compete in a CTF challenge has been heightened after DARPA organized a Cyber Grand Challenge Event in 2016 in Las Vegas \cite{CyberGrand}. In this simplified CTF-like contest, artificial agents were given the possibility to interact with a system exposing a limited number of commands.  

However, interest in the problem of modelling and solving hacking or penetration problems predates this event. Different formalizations of CTF-like problems or penetration testing have been suggested in the literature. Standard models relied on formalism from graph theory (e.g., \emph{Markov decision processes} \cite{sarraute2013penetration}), planning (e.g., \emph{classical planning} \cite{boddy2005course}), or game theory (e.g., \emph{Stackelberg games} \cite{speicher2019towards}); a wide spectrum of models with varying degrees of uncertainty and varying degree of structure in the action space is presented in \cite{hoffmann2015simulated}.

Model-free approaches in which the agent is provided with minimal information about the structure of the problem have been recently considered through the adoption of RL \cite{ghanem2020reinforcement,elderman2016adversarial,pettersson2019using,pozdniakov2020smart}. While these works focus on the application of RL to solve specific challenges, in this paper we analyze the problem of how to define in a versatile and consistent way relevant CTF problems for RL. Notice that, in parallel to this work, some of the problems presented in this paper have already been analyzed and solved with simple RL algorithms in \cite{zennaro2020modeling}. This paper, however, reconsiders particular instances of problem tackled in \cite{zennaro2020modeling} in a wider and more formalized perspective, presenting them within a layered framework of levels of abstraction.

\section{Formalization of Web Hacking} \label{sec:Formalization}

In this section, we explore how the ill-defined problem of web hacking may be \emph{formalized} using different types of standard models (Web hacking $\rightarrow$ CTF $\rightarrow$ game $\rightarrow$ RL problem).

\subsection{From web hacking to CTF}
As discussed in Chapter \ref{sec:Background}, real-world web hacking is an extremely complex problem, with vague success conditions and presenting a wide array of possible courses of action, ranging from the exploitation of publicly known vulnerabilities to reliance on non-technical side-channels like social engineering. 

CTF challenges represent a way to specify web hacking problems. CTFs offer a clear, and yet realistic, way to define web hacking challenges. There are two important advantages in the modelling of web hacking as CTF: (i) CTF challenges have a well-defined objective, and unambiguous termination conditions (either in terms of flag retrieval or time expiration); and, (ii) CTF challenges define an initial restriction on the actions that can be undertaken by a participant (normally requiring all attempts and attacks to take place in the digital domain). 

In this sense we can understand CTFs as a first step in the formalization of web hacking. However, this formalization is still too loose to be useful for machine learning; most importantly, the space of actions, while being implicitly defined, is still too unconstrained to be useful.

\subsection{From CTF to game}
To further our modelling, we can express CTFs in game-theoretic terms. Web hacking CTFs can be defined as a \emph{game}:
\[
\mathscr{G} = \left\langle \mathcal{P},\mathcal{A},\boldsymbol{u}\right\rangle,
\]
where,
\begin{itemize}
    \item $\mathcal{P}$ is a set of players,
    \item $\mathcal{A}$ is a set of actions available to players,
    \item $\boldsymbol{u}$ is a vector of utility or payoff functions, such that $u_i$ is the utility function for player $i$, $1 \leq i \leq \left|\mathcal{P}\right|$.
\end{itemize}

The simplest instance of CTF is a 2-player game with $ \left|\mathcal{P}\right|=2$, where one player is the attacker and the second player is the webserver. As long as the web CTF challenge is static, the webserver may be conceived as a player deterministically reacting to the actions of the attacker. 
As explained in Section \ref{ssec:CTF}, this basic CTF setup may be extended to \emph{adversarial multiplayer games} with $\left|\mathcal{P}\right|=N$, where players are partitioned in a red team and a blue team. In the following, we will focus our attention and our discussion on the 2-player game, although our considerations apply straightforward to the multiplayer case.

For any player, we assume the set of action $\mathcal{A}$ to be finite or countable, so as to allow an artificial agent to select its actions. Notice that this assumption of finiteness or countability is reasonable as long as a CTF takes place in a digital and discrete domain.

The utility function $u_i$ of a player allows for the encoding of the victory condition expressed by a CTF challenge. A stark binary utility function allows to assign a positive utility to the capture of the flag, and a null utility to everything else. More refined utility functions may allow to shape the behaviour of a learned agent more subtly.

A game-theoretic formalization can then be seen as a further step in the process of formalization of web hacking problems. The main contribution in this form modelling, contrasted with a generic CTF model, is the definition of an enumerable set $\mathcal{A}$ of possible actions. This provides the foundation for an agent to choose actions and learn its own action policy. Although game theory already provides tools to analyze web hacking as we have modeled it, this formalization is still not ideal as the modeling of a webserver as an active player results over-generic. In the case of interest, in which we have a single attacker targeting a static system, it would be more practical to describe the webserver as a static component of the game.

\subsection{From game to RL problem}
In the case of web hacking with a static system, the game-theoretic modelling over-defines the webserver by describing it as a player. Alternatively, we can model the game as a \emph{RL problem}:
\[\mathscr{R} = \left\langle \mathcal{S},\mathcal{A},\mathcal{T},\mathcal{R}\right\rangle,\]
where
\begin{itemize}
    \item $\mathcal{S}$ is a set of states the game may be in,
    \item $\mathcal{A}$ is a set of actions,
    \item $\mathcal{T}: \mathcal{S} \times \mathcal{A} \rightarrow \mathcal{S}$ is a state transition function defining how states evolve given an initial state and an action,
    \item $\mathcal{R}: \mathcal{S} \times \mathcal{A} \rightarrow \mathbb{R}$ is a reward function defining the reward obtained by an agent after taking an action in a given state.
\end{itemize}

A RL problem thus defined implicitly assumes a single player. In this model, the webserver is not represented as a second player, but its internal logic is implemented in the state transition function $\mathcal{T}$. The state transition function specifies how the system reacts upon the action of the playing agent, and its dynamics relies on two assumptions.
First, we assumed that, in general, the result of an action $a \in \mathcal{A}$ depends not only on the action itself, but also on the current state $s\in \mathcal{S}$ of the system. This correspond to the assumption of a \emph{stateful} system. This assumption is meaningful, as real web systems may be in different states after interacting with their users. Notice that a stateless system can, in any way, be considered as a limit case of a stateful system with a single unchanging state.
Second, we assumed that, in general, the result of an action $a \in \mathcal{A}$, given the current state $s\in \mathcal{S}$, may be \emph{stochastic}. This assumption is meaningful in that real web systems may rely on stochastic functions. Moreover, such an assumption may allow us to model potential network communication fails or attempts by the system to obfuscate its logic. Notice that a deterministic state can, in any way, be considered as a limit case of a stochastic system with a delta distribution function.
In sum, we express the logic of the webserver as a probabilistic transition function $\mathcal{T} = P\left( s' \vert s, a \right)$ specifying a probability distribution over future states $s'$, given the current state $s$ and action $a$. We will refer to $\mathcal{T}$ as the transition function, the logic of the game, or the dynamics of the environment.

As in the game-theoretic formulation, the set of action $\mathcal{A}$ is a countable set of actions available to the agent.

The reward function $\mathcal{R}$ translates the utility funciton $\boldsymbol{u}$ from the game-theoretic modelling to the RL formalism.

Finally, the set of states $\mathcal{S}$ allows for the tracking of the state of the game. Notice that, although the state of the game is uniquely determined at any point in time, the agent may not be aware of it. This leads to a \emph{partially observable} game, in which the agent has no certain knowledge about the current state of the system, but it has only belief over the possible states. Through its own local state, which encodes its imperfect knowledge, the agent tries to keep track of the actual state of the system. Notice that a completely observable game may be considered as a limit case in which all the beliefs collapse into delta functions.

This final RL formalization captures well enough the problem of web hacking: it is flexible enough to accommodate very different hacking challenges, but, at the same time, is constrained enough that all its component are well-defined so that standard RL algorithms may be used to train artificial agents. We will then make the RL formalization the theoretical blueprint of our model for web hacking.

\section{The Agent Web Model} \label{sec:AgentWebModel}

In this section we use the RL formalism defined in Section \ref{sec:Formalization} to characterize our own model for web hacking. We then discuss how this generic model may be used to implement actual web hacking problems at different levels of abstraction.

\subsection{The Agent Web Model}
In order to define a RL problem, it is necessary to define the state transition function of the problem. In our context, this function represents the logic of the target webserver. Different systems, with different types of vulnerabilities, may be represented in different ways.
To simplify the modelling of a webserver, we will represent it as a collection of generic \emph{objects}. These objects are taken to represent entities of interest (e.g.: files, ports) that can be targeted by the actions $\mathcal{A}$ of an attacker. This simplification allows us to decompose the design of a target system, its logic and its states. Transition functions can be defined in a modular way with respect to specific objects, and the state of the system may be factored in the state of single objects.

The decomposition of a webserver into a collection of objects also allows us to easily define instances of webservers at different \emph{levels of abstraction}. By defining the nature and the number of existing objects, and by defining which actions an agent can take in relation to the defined objects, we can immediately control the complexity of the RL problem at hand.

Moreover, another aim of ours in having a modular system defined in terms of individual objects, is the possibility of instantiating new challenges in an automatic, possibly random, way. Such a \emph{generative model} of web-hacking problems would provide the opportunity to easily generate a large number of problems on which to train a RL agent.

We call this flexible, generative model to instantiate different types of web hacking problems, the \emph{Agent Web Model}.

\subsection{Levels of abstraction}

\begin{figure}
\centerline{\includegraphics[scale=.7]{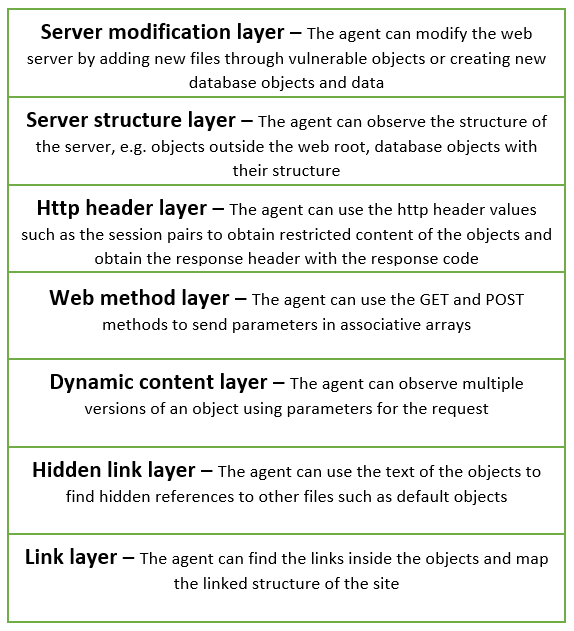}}
\caption{Levels of abstraction in the Agent Web Model.}
\label{fig:ModelOverview}
\end{figure}

Concretely, we define 7 different levels of abstraction for web hacking with increasing complexity in terms of the actions and the feedback that the agent can receive (see Figure \ref{fig:ModelOverview}). 
We start at level1 with the model of a very simple website, composed of basic files, without web parameters and sessions. At higher levels, we allow the agent to interact with more complex objects making up the website: requests to files can accept multiple input parameters with different web methods, as well as multiple session values. 
A hacking problem at level1 has a trivial solution which could be coded manually in a simple algorithm, but we will show that the computational complexity soon escalates as we move up in the levels. A hacking problem at level7 is close to real-world web hacking, where an attacker can even create its own objects on the target site (e.g. command script) and carry out complex exploitation strategies; this sort of problem is far from a trivial solution. 

In the following, we discuss the details of the different layers of the Agent Web Model, including the number of states and actions that have to be handled in different levels. Except when explicitly stated, in all levels of abstractions we will assume that the objects on a webserver are files, and we will take a simple binary reward function $\mathcal{R}$ that returns a unitary reward when the agent accomplishes its task, and zero otherwise.

\subsubsection{Level1 - Link layer \label{sec:Level1}}
\begin{figure}
\centerline{\includegraphics[scale=.8]{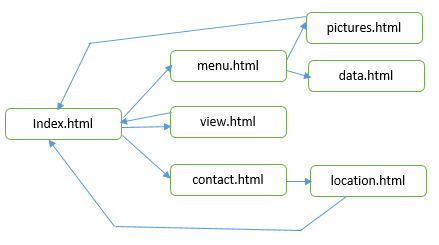}}
\caption{Example of webserver at level1. \newline Nodes represent files and solid arrows represent connections between files.}
\label{fig:Level1}
\end{figure}

In level1, a website is composed by a set $\mathcal{O}=\{ \FILE{1},\allowbreak \FILE{2},\allowbreak ...  ,\allowbreak \FILE{N} \}$ of objects representing simple static HTML files. We take the first file to represent the \textit{index.html} file inside the webroot.  Files are linked to each others by pointers, and one of the files contains the flag. All the files can be accessed by the agent without restrictions; no parameters are required, and the HTTP headers have no meaningful information such as sessions. The actual file content is irrelevant, except for the case of the flag. Practically, level1 problems can be represented as a directed graph of files (see Figure \ref{fig:Level1}).

The set of actions comprises only two parametric actions: $\mathcal{A}=\left\{ \READi, \SEARCHi \right\} $. The action $\READi$ reads the $i^{\textrm{th}}$ file and returns the list of linked files. The action $\SEARCHi$ checks the $i^{\textrm{th}}$ file for the presence of the flag. See Table \ref{Tab:Level1actions} for a summary of the actions, their parameters and their return values. Note that these actions can be performed only on files that the agent has discovered on the remote webserver.

\begin{table}[htbp]
\caption{Actions in level1}
\begin{center}
\begin{tabular}{cccc}
\hline
\textit{Action name} & \textit{Parameters} & \textit{Result} \\
\hline
$\READ{}$ & file & set of files \\
$\SEARCH{}$ & file & true/false \\
\hline
\end{tabular}
\label{Tab:Level1actions}
\end{center}
\end{table}

Without training a RL agent, a simple heuristic solution to this problem would be to read the files one by one in order to discover all files, and then search for the flag inside each one.

The number of files $N$ that a website hosts has a significant influence on the problem scale. The actual size of the action space $\left\vert \mathcal{A} \right\vert$ depends on the value of $N$: an agent can take up to $2N$ different actions, that is, a $\READ{}$ action and a $\SEARCH{}$ action for each file. Moreover, an agent is required to keep track of its own knowledge state, that is record what actions has been executed and what result was observed. A basic agent can simply track, for each file, whether action $\READ{}$ was tried ($2^{N}$ states) and whether action $\SEARCH{}$ was tried ($2^{N}$ states). In total, it will have \textit{$2^{2N-1}$} states; Table \ref{Tab:Level1states} shows an estimate of the number of actions and states as a function of the number of files. 

\begin{table}[htbp]
\caption{Number of actions and states in level1}
\begin{center}
\begin{tabular}{cccc}
\hline
\textit{Number of files} & \textit{Number of actions} & \textit{Number of states} \\
\hline
$N$ & $2N$ & $2^{2N-1}$ \\
2 & 4 & 8 \\
3 & 6 & 32 \\
5 & 10 & 512 \\
10 & 20 & $\approx 5 \cdot 10^5$  \\
\hline
\end{tabular}
\label{Tab:Level1states}
\end{center}
\end{table}

\subsubsection{Level2 - Hidden link layer\label{sec:Level2}}
\begin{figure}
\centerline{\includegraphics[scale=.8]{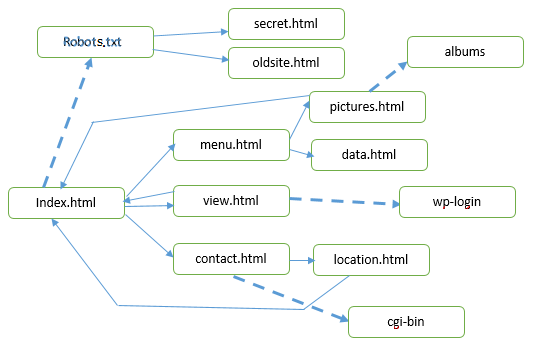}}
\caption{Example of webserver at level2
\newline Nodes represent files, solid arrows represent direct connections, dashed arrows represent indirect connections between files.}
\label{fig:Level2}
\end{figure}

In level2, we model again the website as a collection of static HTML files. Files are still linked by pointers, but we now distinguish two types of pointers: links that are openly visible to the attacker upon reading the files (as it was in level1), and implicit pointers that requires an actual analysis of the file. Real-world examples of these types of implicit pointers may be: comments in the source code that refers to another file without stating a direct link; keywords used in the file that refer to a special type or version of a webserver app or CMS, and that indicate the existence of other default files; recurrent appearance of a word, suggesting that there may be a file or folder with the same name.
Practically, level2 problems can be represented as directed typed graph of files with two types of edges (see Figure \ref{fig:Level2}). 

The set of actions of the agent is composed now of three parametric actions $\mathcal{A}=\left\{\mathtt{read}\middlebreak(\textrm{file}_{i}),\middlebreak \mathtt{search}\middlebreak(\textrm{file}_{i}),\middlebreak \mathtt{deepread}\middlebreak(\textrm{file}_{i})\right\}$. As before, action $\READi$ read the $i^{th}$ file and returns a list of files connected by an \emph{explicit} link, while $\SEARCHi$ checks the $i^{th}$ file for the presence of the flag. The action $\DEEPREADii$ processes the $i^{th}$ file and returns a list of files connected by \emph{implicit} links. See Table \ref{Tab:Level2actions} for a summary of the actions, their parameters, and their return values.
Notice that, at this level of abstraction, the logic and the algorithm for performing a $\DEEPREAD{}$ is implicitly provided in the game itself. At higher levels of abstraction, the task of actually parsing a HTML file and uncover the possible URLs of new files would be delegated to the learning agent; such an agent would receive the actual content of a file and it could use a range of algorithms to process the text, from simple dictionary mapping (e.g.: \textit{apache} mapping to \textit{cgi-bin}, \textit{wordpress} mapping to \textit{wp-login}, etc.) to more complex natural language processing neural networks able to propose new potential file candidates.   

\begin{table}[htbp]
\caption{Actions in level2}
\begin{center}
\begin{tabular}{cccc}
\hline
\textit{Action name} & \textit{Parameters} & \textit{Result} \\
\hline
$\READ{}$ & file & set of files \\
$\DEEPREAD{}$ & file & set of files \\
$\SEARCH{}$ & filename & true/false \\
\hline
\end{tabular}
\label{Tab:Level2actions}
\end{center}
\end{table}

Given $N$ files on the webserver, the cardinality of the action space is now $\left\vert A \right\vert = 3N$ and the cardinality of agent state space is $2^{3N-1}$, by trivially scaling up from level1 because of an additional action. Table \ref{Tab:Level2states} shows estimates for few values of $N$.

\begin{table}[htbp]
\caption{Number of actions and states in level2}
\begin{center}
\begin{tabular}{cccc}
\hline
\textit{Number of files} & \textit{Number of actions} & \textit{Number of states} \\
\hline
n & $3N$ & $2^{3N-1}$ \\
2 & 6 & 32 \\
3 & 9 & 256 \\
5 & 15 & 16384 \\
10 & 30 & $\approx 5.3\cdot 10^8$  \\
\hline
\end{tabular}
\label{Tab:Level2states}
\end{center}
\end{table}


\subsubsection{Level3 - Dynamic content layer\label{sec:Level3}}
\begin{figure}
\centerline{\includegraphics[scale=.6]{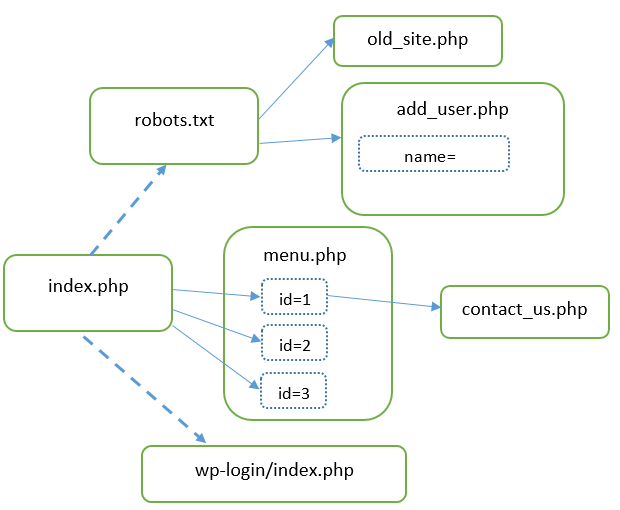}}
\caption{Example of webserver at level3
\newline Solid nodes represent files, dotted nodes within a file illustrate a pair of parameter name and value that may be sent to a file, solid arrows and dashed arrows represent respectively direct and indirect connections between files given a parameter pair.
\newline If an arrow leads to a file, it means that upon a successful $\READ{}$ or $\DEEPREAD{}$ action the file itself is revealed without parameters; if an arrow leads to an internal dotted node, then after a successful $\READ{}$ or $\DEEPREAD{}$, a file together with a parameter list for the file is also sent back to the agent.
}
\label{fig:Level3}
\end{figure}

The real complexity of a website starts with server-side scripting. In level3 we consider a webserver that can dynamically execute server-side scripts by processing user parameters and generating static content for the client. A single web file can provide multiple results based on the parameters that the site receives from the client. 
We still model the webserver as a collection of static files, delegating the complexity of dynamic server-side scripting in the space of actions. From a practical perspective, the webserver can still be seen as directed typed graph with nodes that may return different values depending on the received parameter (see Figure \ref{fig:Level3}).

In order to account for parameter passing, we now define a new set of parametric actions: $\mathcal{A}=\left\{ \READiii,\middlebreak \SEARCHiii,\middlebreak \DEEPREADiii \right\}$. Actions have the same semantics as in level2, but now, beyond receiving file $i$ as an input parameter, they also receive parameter name $j$ and parameter value $k$. This reflects the request of a specific URL (file $i$) together with a specific parameter (parameter name $j$) and a set value (parameter value $k$). 
The return value of the $\READ{}$ and $\DEEPREAD{}$ actions is also enriched by a possible set of parameter names and values; this is due to the fact that the answer of the webserver may contain not only links to other files, but it may include the specific parameter pairs relevant to the connected files.
See Table \ref{Tab:Level3actions} for a summary of the actions, their parameters, and their return values. 
Notice that at this level of abstraction, we assume that only a single pair $(\PNAME{j}, \PVAL{k})$ can be specified as input; moreover, to keep the complexity in check, we assume that $\PNAME{j}$ and $\PVAL{k}$ may assume values in a finite set, that is $1 \leq j \leq M$ and $1 \leq k \leq O$, $M,O \in \mathbb{N}_{\geq 0}$.

\begin{table}[htbp]
\caption{Actions in level3}
\begin{center}
\begin{tabular}{cccc}
\hline
\textit{Action name} & \textit{Parameters} & \textit{Result} \\
\hline
$\READ{}$ & file, & set of files,\\
& parameter name, & set of parameter  \\
& parameter value& names and values \\
$\DEEPREAD{}$ & file, & set of files,\\
& parameter name, & set of parameter  \\
& parameter value& names and values \\
$\SEARCH{}$ & file, & true/false \\
& parameter name, &  \\
& parameter value & \\
\hline
\end{tabular}
\label{Tab:Level3actions}
\end{center}
\end{table}

The cardinality $\left\vert \mathcal{A} \right\vert$ of the action space is now much larger because of combinatorial explosion in the parameters of an action. Assuming $N$ files on the webserver, and a set of $M$ parameter names and $O$ parameter values that can be freely combined, each action can be instantiated $N + NMO$ times ($N$ times without parameters, and $NMO$ times considering all combinations). In total, we then have $3(N+NMO)$ concrete actions the agent can take. A trivial agent that explicitly tracks its state by recording, for each possible action, if it was taken or not, would have to deal with a space with a cardinality of $2^{3(N+NMO)-1}$. Table \ref{Tab:Level3states} shows some estimates for different values of $N$, $M$, and $O$.

\begin{table}[htbp]
\caption{Number of actions and states in level3}
\begin{center}
\begin{tabular}{ccccc}
\hline
\textit{\#files} & \textit{\#pars} & \textit{\#pvals} &\textit{\#actions} & \textit{\#states} \\
\hline
N & M & O & $3(N+NMO)$ & $2^{3(N+NMO)-1}$ \\
2 & 2 & 2 & 30 & $\approx 5.4\cdot10^{8}$ \\
2 & 5 & 5 & 156 & $\approx4.6\cdot10^{46}$ \\
5 & 2 & 2 & 75 & $\approx1.9\cdot10^{22}$ \\
5 & 5 & 5 & 390 & $\approx1.3\cdot10^{117}$  \\
10 & 5 & 5 & 780 & $\approx3.2 \cdot 10^{234}$  \\
\hline
\end{tabular}
\label{Tab:Level3states}
\end{center}
\end{table}

\subsubsection{Level4 - Web method layer\label{sec:Level4}}
\begin{figure}
\centerline{\includegraphics[scale=.55]{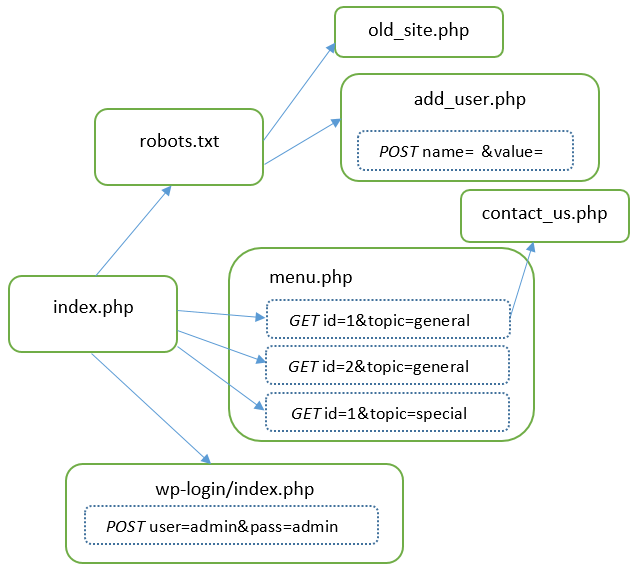}}
\caption{Example of webserver at level4
\newline Solid nodes represent files, dotted nodes within a file illustrate possible lists of parameter names and values that may be sent to a file via a webmethod, solid arrows represent connections between files given parameters. Inside the internal nodes the used webmethod is indicated.}
\label{fig:Level4}
\end{figure}

In level4 we scale the complexity in an effort to make the problem more realistic. We now consider the possibility of a webserver receiving a request specifying a web method and containing a list of parameter names and an associated list of parameter values. This better capture the actual dynamics of the HTTP protocol, reflecting the syntax of common HTTP methods such as \textit{GET} and \textit{POST}. The webserver is always modeled as a collection of files forming a directed typed graph with nested nodes (see Figure \ref{fig:Level4}). 

The set of parametric actions is now restructured. We drop the previous artificial distinction between $\READ{}$, $\DEEPREAD{}$; while in previous levels of abstraction the task of extracting explicit and implicit links was externalized in the logic of the game, from now on it is the task of the agent to parse and analyze the answer of the webserver in order to find explicit and implicit links, as well as the flag itself. The new action set is: $\mathcal{A}=\left\{ \GETiv, \middlebreak \POSTiv\right\}$. The $\GET{}$ and $\POST{}$ actions implement the respective web methods, and they receive as input a file ($\FILE{i}$), a list of parameter names ($[\PNAME{}]$) together with a list of parameter values ($[\PVAL{}]$). The result of these actions is a HTTP page. The flag is considered retrieved when the agent obtains the HTTP page containing the flag. See Table \ref{Tab:Level4actions} for a summary of the actions, their parameters, and their return values.


\begin{table}[htbp]
\caption{Actions in level4}
\begin{center}
\begin{tabular}{ccc}
\hline
\textit{Action name} & \textit{Parameters} & \textit{Result} \\
\hline
$\GET{}$ & file, & HTTP page\\
& set of parameter names, & \\
& set of parameter values &  \\
$\POST{}$ & file, & HTTP page\\
& set of parameter names, & \\
& set of parameter values &  \\
\hline
\end{tabular}
\label{Tab:Level4actions}
\end{center}
\end{table}

Given, as before, $N$ files on the webserver, $M$ possible alternatives for the parameter names, $O$ possible alternatives for the parameter values, the cardinality $\left\vert \mathcal{A} \right\vert$ depends on the maximum length $P$ of the list of parameters. With $P=0$, $\left\vert \mathcal{A} \right\vert = 2N$, that is, trivially, $\GET{}$ and $\POST{}$ actions with no parameter on each file. With $P=1$, $\left\vert \mathcal{A} \right\vert = 2N + 2NMO$, that is, the same two actions for every possible combination of zero or one parameter name and value (similar to level3). In the worst case in which $P=M$, that is the list can be long enough to contain all the parameter names, the number of possible actions could be estimated as:

\resizebox{0.8\hsize}{!}{%
\[
\underbrace{2}_{\begin{array}{c}
actions\\
{}
\end{array}} \cdot
\underbrace{N}_{\begin{array}{c}
files\\
{}
\end{array}} \cdot
\underbrace{\sum_{i=0}^{M}}_{\begin{array}{c}
all\,list\\
lenghts
\end{array}} \cdot 
\underbrace{\binom{M}{i}}_{\begin{array}{c}
all\,combinations\\
of\,i\,param\,names
\end{array}} \cdot 
\underbrace{{O}^{i}}_{\begin{array}{c}
all\,combinations\\
of\,i\,param\,vals
\end{array}}
\]}

\normalsize
A trivial agent that would store again its state knowledge about actions using binary values would have to deal with a state space of cardinality $2^{\left\vert \mathcal{A} \right\vert}$.




\subsubsection{Level5 - HTTP header layer\label{sec:Level5}} 
While all the previous layers considered only the URL and the body part of the HTTP packets, level5 takes the HTTP header into consideration as well. The HTTP header can contain relevant information such as the session variables  or the web response code in the response header. The session, which is composed by a session variable name and value (e.g., \textit{JSESSIONID=Abvhj67}), is used to provide elevated access to special users; a practical example is the login process (which may happen by sending multiple \textit{POST} parameters, as modeled in level4), after which the server sets a new session value. Additional HTTP header information, such as the browser type or the character encoding, can also have an effect on the response provided by the webserver.

\begin{figure}
\centerline{\includegraphics[scale=.6]{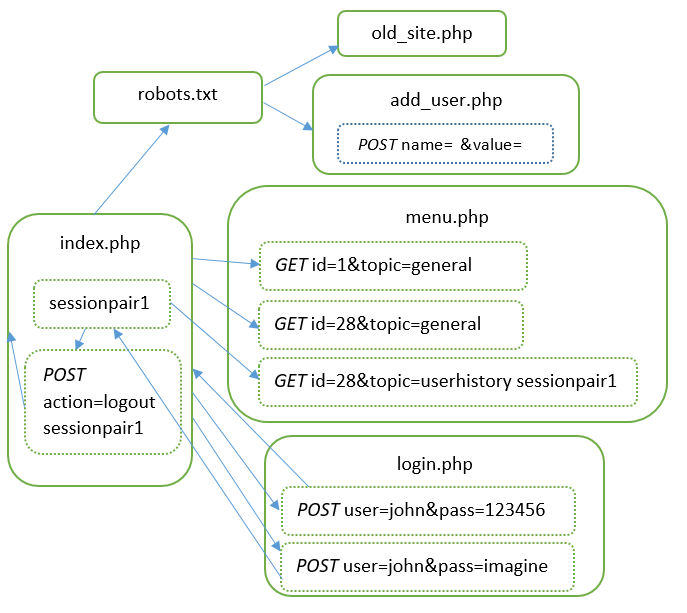}}
\caption{Example of webserver at level5
\newline Solid nodes represent files, dotted nodes within a file illustrate possible lists of parameter name and value pairs and session name and value pairs that may be sent to a file via a webmethod, solid arrows represent respectively connections between files given parameters and sessions.}
\label{fig:Level5}
\end{figure}

We always model the webserver as a collection of files forming a directed typed graph with nested objects (see Figure \ref{fig:Level5}). Object access is now more complex as it depends also on all the header variables.
This complexity may in part be reduced by considering a pair of session name and session value as a single parameter (session values are usually random numbers with high entropy so there is no point in handling the session variable name and value separately unless the session values are predictable and the attacker wants to brute-force the session value), and by limiting the number of allowed session pairs and HTTP headers. 
Under this simplification, we preserve the same actions as level4, but we extend the signature of their input parameters: $\mathcal{A}=\left\{ \GETv, \allowbreak \POSTv\right\}$. Beside receiving an input file ($\FILE{i}$), a list of parameter names ($[\PNAME{}]$) together with a list of parameter values ($[\PVAL{}]$), the $\GET{}$ and $\POST{}$ methods now also receive a list of session pairs ($[\SPAIR{}]$) and a HTTP header ($\HTTPHEADER{}$). The result of these actions is a web response, possibly together with a HTTP page. The web response code (e.g., 200, 404, 500) reflects the accessibility of the requested object. As before, the flag is considered retrieved when the agent obtains the HTTP page containing the flag.
See Table \ref{Tab:Level5actions} for a summary of the actions, their parameters, and their return values.

With reference to the actions we have defined, we observe an enlargement of the action space that now depends on the number $N$ of files on the server, the number $M$ of parameter names that can selected, the number $O$ of parameter values available, the number $P$ of parameter pairs that can be sent, the number $Q$ of session pair values available, the number $R$ of session pairs that can be sent, and the number $S$ of HTTP header without cookies that can be sent.

Figure \ref{fig:Level5} provides also the illustration of a possible interaction between the agent and the webserver. The attacker first tries to log in using an invalid password, which actually reveals a new version of the \textit{login.php} file by redirecting the page to the \textit{index.php} page without session. Using the right credentials shows another version of the \textit{login.php} page that instead redirects the user to a version of \textit{index.php} with the session pair \textit{sessionpair1}. This version of the \textit{index.php} leads then to another version of the file (logout action) that is connected to the original version of \textit{index.php} without session.

\begin{table}[htbp]
\caption{Actions in level5}
\begin{center}
\begin{tabular}{ccc}
\hline
\textit{Action name} & \textit{Parameters} & \textit{Result} \\
\hline
$\GET{}$ & file, & HTTP page,\\
& set of parameter names, & web response   \\
& set of parameter values & \\
& set of session pairs  &  \\
& HTTP header & \\
$\POST{}$ & file, & HTTP page,\\
& set of parameter names, & web response   \\
& set of parameter values & \\
& set of session pairs  &  \\
& HTTP header & \\
\hline
\end{tabular}
\label{Tab:Level5actions}
\end{center}
\end{table}




\subsubsection{Level6 - Server structure layer\label{sec:Level6}} 

\begin{figure}
\centerline{\includegraphics[scale=.55]{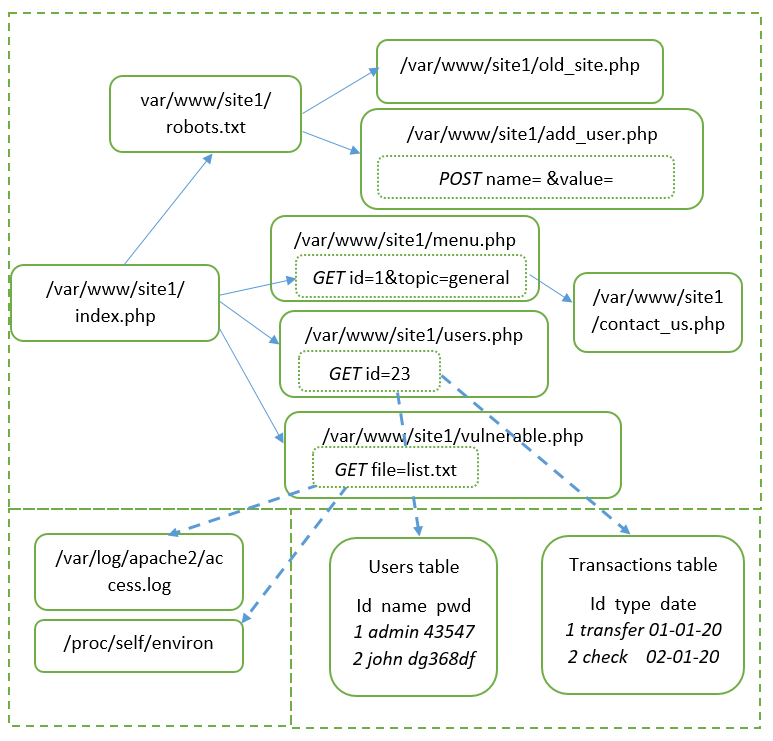}}
\caption{Example of webserver at level6
\newline 
Solid nodes represent files, dotted nodes within a file illustrate possible lists of parameter name and value pairs and session name and value pairs that may be sent to a file via a webmethod, solid arrows represent connections between files given parameters and sessions. Dotted boundary lines separate different logical spaces, such as the webserver space and the database space. Dashed arrows mark connections between these logical spaces.}
\label{fig:Level6}
\end{figure}

In a complex web hacking scenario, the attacker may map the file system of the server in order to collect information to be used during the attack. In level6 we extend the formalization of the webserver in order to consider not only files within the webroot, but also objects beyond it, such as local files and databases. This extension allows to simulate attacks relying on local file inclusion (LFI) vulnerabilities, or information gathering attacks on a database in order to set up a SQL injection. Figure \ref{fig:Level6} shows the structure of a webserver, and it illustrate a possible LFI attack to obtain the webserver logs or the environmental variables.

While the action set remains the same as level5, the extension of the domain of the objects beyond the webroot escalates the number of targets that the agent may consider. Complexity soars with the increase of objects, including databases, and, within a database, its tables, columns and rows.

Level6 abstraction provides the agent the following additional features compared to lower level of abstractions:
\begin{itemize}
    \item Obtaining the local resources of the website such as the background files if there is any or the background database records used for the website operation. The attacker can use these data for the attack with the normal requests covered in lower layers;
    \item Accessing the data in order to compromise other websites residing on the same webserver;
    \item Obtaining the webserver files that are used for other purposes than the website operations, such as users data, other service data, operating sytem data and use these for the attack.
\end{itemize}

In this scenario the access rights of the objects play an important role; running a webserver as a root can have serious consequences, while having minimum access rights reduce the chance of such exploitations. Notice, though, that practically, from the point of view of the agent, there is no difference between the cases when an object is not present or the object is present but there is no read access for the object by the website. 


\subsubsection{Level7 - Server modification layer\label{sec:Level7}} 

The last layer we consider in our Agent Web Model is the server modification layer. In this layer we assume that the agent can carry out complex meaningful web hacking actions such as creating its own objects, either inside or outside the web root. With the ability to create its own files, the attacker can place command scripts that can be used to carry out advanced attacks. Figure \ref{fig:Level6} show the same structure of the server as in level6, and it illustrates an attacker creating its own files on the webserver.

Attacking actions leading to the creation of objects can be carried out by the web requests that we have already considered. The action does not change, but the domain of the parameters increases in order to allow for more sophisticated actions.

\begin{figure}
\centerline{\includegraphics[scale=.55]{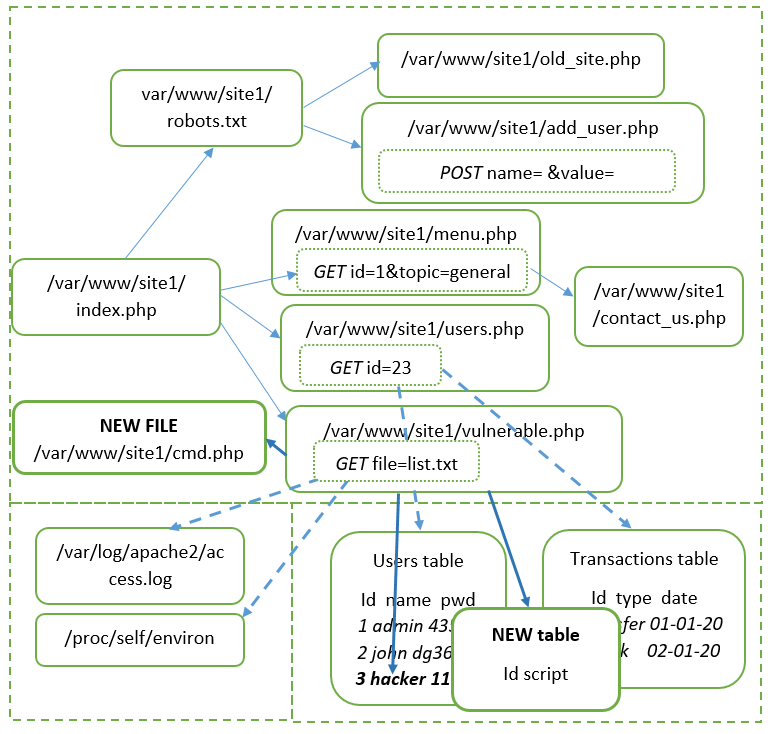}}
\caption{Example of webserver at level7
\newline Solid nodes represent files, dotted nodes within a file illustrate possible lists of parameter name and value pairs and session name and value pairs that may be sent to a file via a webmethod, solid arrows represent  connections between files given parameters and sessions. Dotted boundary lines separate different logical spaces, such as the webserver space and the database space. Dashed arrows mark connections between these logical spaces. Boldface objects represent objects created by the attacker.}
\label{fig:Level7}
\end{figure}

Level7 abstraction provides the agent the following additional features compared to lower level of abstractions:
\begin{itemize}
    \item Causing denial of service by editing important objects for the site operation;
    \item Defacing the site by changing the site content;
    \item Escalating privileges by adding data to objects;
    \item Uploading attack scripts to provide extra functions for the attack;
    \item Removing attack clues by deleting log files, deleting temporary files that were used for the attack.
\end{itemize}

Level7 is assumed to be the highest level of modelling, capturing all relevant features of hacking; thus, solving this challenge is extremely hard, and we would expect that a successful agent would perform as well as, or better than, a professional human hacker actually involved in a process of website hacking.

\section{Modelling web vulnerabilities} \label{sec:ModellingWeb}
In this section we analyze how different types of web vulnerabilities fit within our \emph{Agent Web Model}. Refer to Table \ref{tab:Vulnerabilities_in_the_model} for a summary of which vulnerabilities may be modeled at which level.

\textbf{Information disclosure} is a type of vulnerability where the attacker gains useful information by penetrating the system. Evaluating the usefulness of the gained information is not trivial, but through the CTF formalization we make the simplifying assumption that relevant information the attacker may be interested into is marked by the flag. In this way, it is possible to equate successful information disclosure with the retrieval of the flag. Every level of abstraction in our \emph{Agent Web Model} captures this attack: in level1 sensitive information (flag) is in a public linked file on the webserver; in level2 sensitive information (flag) can be inside a private file; in the following layers (level3 to level5) sensitive information (flag) can be accessed using special parameters or sessions; in level 6, sensitive information (flag) can be inside a file outside the webroot.    

\textbf{Web parameter tampering} \cite{24sins4} is a type of attack where the web parameters exchanged by the client and the server are modified in order to have access to additional objects. Our \emph{Agent Web Model} captures this attack starting at level3 by allowing the specification of web parameters in the URL; in level4 it is possible to add HTTP body parameters (POST message); in level5 it is possible to edit cookies in the HTTP header. In all these instances, an agent can perform web parameter tampering either by meaningfully exploring the space of possible values of these parameters, or by trying to brute-force them.

\textbf{Cross Site Scripting (XSS)} attacks \cite{XSSAttack} enable attackers to inject client-side (e.g. javascript) code into the webpage viewed by other users. By exploiting a XSS vulnerability the attacker can overwrite the page content on the client side, redirect the page to the attacker's page, or steal the valid sessions inside the cookie. All these offensive actions can be followed by some social engineering trick in case of a real attack. In the context of CTF style challenges where additional clients are not available, the aim of an attacker is simply to show the existence of the vulnerability. A flag may be used to denote a page that is only accessible indirectly by redirection. The task for the agent is to find the right parameters to achieve the redirection. The injected client-side code for XSS has to be sent through web parameters. XSS attacks can be simulated in our \emph{Agent Web Model} as soon as we can interact with parameters: in level3 the attacker may add code in the URL; in level4 the attacker may modify POST parameters; in level5 the XSS attack may affect the header. 

\textbf{Cross Site Request Forgery (CSRF)} \cite{siddiqui2011csrf} is a type of vulnerability where the attacker sends a link to authenticated users in order to trick them to execute web requests by social engineering. If the users are authenticated (have sessions) the malicious request (e.g., transferring money, changing the state) is  executed by the server.
This exploitation is based on social engineering and on misleading the user. In addition, CSRF tokens are sent by the server to filter out unintended requests; the agent can check the existence of appropriate CSRF tokens or exploit requests with weak CSRF tokens.
In our model the CSRF attack has to be simplified to consider only the CSRF token manipulation in layer 5.  

\textbf{SQL injection} \cite{advancedsqli} is a vulnerability where malicious SQL statements can be executed by the server due to the lack of input validation on the server side. By modifying the original SQL statement of a server-side script the attackers can bypass authentication, access confidential database information or even write attack scripts on the server (select into outfile command). In most of the cases the attacker has to map the database structure of the target by finding, for instance, the different table names along with their column names and types.
In our \emph{Agent Web Model} this attack can be completely simulated at level6, although other simplified versions may happen at lower levels. In the easiest case the agent only need one dynamic parameter without sessions; bypassing only a simple authentication or collecting data from the same table that the server-side script uses does not require to know the table name and other database structure data; in these cases, a basic form of SQL injection may be simulated even in level3. Complex cases comprising all the database parameters need to happen at level6. If the attacker uses the SQL injection to carry out further actions such as writing attacking scripts on the compromised site, then this has to happen at level7. 
All the above mentioned cases require a very high number of actions especially when the agent has to execute a Boolean-based blind SQL injection. In these cases, the vulnerable application provides only true or false answers, so obtaining one single piece of information, such as a column name in a table, requires binary search type requests for each character, which can lead to an exponential number of actions.
Notice that the Agent Web Model abstraction does not consider the response time of the environment. In very specific cases such as time-based blind SQL injections, the attacker may have to measure the response time; this type of exploitation would require the consideration of the server reaction time too. 

\textbf{Xpath injection} \cite{Xpath} is a web vulnerability where the attacker injects code into the web request, but the target of the attack is not a database (as in the case of SQL injection) but an XML file. By exploiting Xpath injection the attacker can iterate through XML elements and obtain the properties of the nodes one by one. This operation requires only one parameter, so simulating Xpath injection is theoretically possible in level3. Since the exploitation of the Xpath injection does not require the name of the XML file, mapping the files outside the webroot is not necessary even if the XML file is outside the webroot. On the other hand, the vulnerable parameter can be a POST parameter (level4) or it can require a specific session (level5). 

\textbf{Server-Side Template Injection (SSTI)} \cite{SSTI} is a vulnerability where the attacker uses native template syntax to inject a malicious payload into a website template. For the exploitation the agent has to use additional actions that are SSTI specific, such as sending a string like \$\{7*7\} together with a parameter. Theoretically, an easy SSTI vulnerability can be exploited in level3, but all other layers above can be used to represent specific attack cases (vulnerable parameter in POST on level4, session required for exploitation on level5); in particular cases, the attacker can list the server structure (level6) or can create files with arbitrary code execution (level7).

\textbf{File inclusion} \cite{fileinclusion} makes the attacker capable of including remote or local files by exploiting a vulnerable web parameter on the website. In case of remote file inclusion (RFI), the attacker can include its own remote attacking script in the server-side script. Remote file inclusion can have very serious consequences, but in a CTF challenge the aim is just to show the possibility of the exploitation, not to carry out an actual exploit. RFI can be realized by providing a remote file that sends the flag if the request is initiated from the target website IP. Exploiting RFI is possible in level3 but other parameters, such as POST request and sessions, can be relevant (level4 and level5). As a consequence of the RFI vulnerability the attacker can create files on the website for further attacks.
In case of local file inclusion (LFI), the attacker can include local files in the server-side script. For the exploitation one single parameter is theoretically enough, but since usually it is necessary to read local files outside the webroot, the agent has to map at least a part of the server structure (level6). In some exploitation scenarios the attacker can use local files (such as logs or files in the \emph{/proc} linux folder) to create its own files on the server (level7).

\textbf{Session-related attacks} \cite{SessionRelated} exploit session disclosure or other weaknesses in the session generation process. Since we model the environment as the server itself without other network nodes, man in the middle session disclosures cannot be considered. Other session disclosures can be possible, for instance, if the sessions are stored in the logs and the website can access the log files (LFI), as modeled on level6. Brute-forcing the session is also possible in level5, but brute-force actions increase dramatically the complexity and the number of possible actions.

\textbf{HTTP response splitting} \cite{24sins2} is a vulnerability where the attacker can control the content of the HTTP header of a web request. The ability of the attacker to construct arbitrary HTTP responses can result in many other exploits such as cache poisoning or Cross Site Scripting. Our \emph{Agent Web Model} considers the HTTP header information in level5, but only with limited information (different session pairs and the whole header together with different versions). Training the agent to learn HTTP response splitting exploitation would require to split the HTTP header in multiple parts and allow the agent to consider actions on different HTTP header combinations.

\begin{table}[htbp]
\caption{Web vulnerabilities in the Agent Web Model}\label{tab:Vulnerabilities_in_the_model}
\begin{center}
\begin{tabular}{ccc}
\hline
 & \textit{Agent} &  \\
\textit{Web} & \textit{Web} & \textit{Comment}\\
\textit{Vulnerability}&\textit{Model}&\\
&\textit{Layers}&\\
\hline

Information & 1-6 & Flag in web files,\\ 
disclosure & &  header or outside object\\
\hline
Parameter & 3-5 & Flag access with modified\\
tampering & & parameters or sessions \\
\hline
Cross Site & 3-5 & Flag access with\\
Scripting & & client-side redirection\\
\hline
Cross Site & 5 & Only CSRF token\\
Request Forgery & & manipulation \\
\hline
SQL injection & 3-7 & Flag with authentication\\
&& bypass or from database\\
\hline
Xpath injection & 3-4 & Flag with authentication\\
&& bypass or from XML file\\
\hline
Server-Side & 3-7 & Flag with object access\\ 
Template Injection && or privilege escalation\\
\hline
Remote File & 3 & Flag with remote \\
Inclusion && file access\\
\hline
Local File & 3-7 & Flag with local\\
Inclusion && file access\\
\hline
Session related & 5-6 & Flag with session\\
attacks && manipulation \\
\hline
HTTP response & 5 & Flag with HTTP header\\
splitting && manipulation\\
\hline
\end{tabular}
\end{center}
\end{table}

\section{Implementation of the Agent Web Model} \label{sec:Implementation}
An implementation of the first three levels of the Agent Web Model has been developed in agreement with the standard defined in the \emph{OpenAI gym} framework \cite{brockman2016openai}, and it has been made available online\footnote{\url{https://github.com/FMZennaro/gym-agentwebmodel}}. Each level provides a simple interface to an abstraction of a CTF challenge; a webserver makes available to the agent a finite set of actions, and it returns information about the state of the game upon the choice of an action. Levels may be instantiated parametrically (deciding the number of files, the links, and the possible parameters), thus offering the possibility of generating a wide variety of challenges for a learning agent. The first three levels already offer a wide degree of challenge: while level1 provides a simple, tutorial-like, CTF game, the third level constitute a non-trivial abstraction of a real hacking challenge.

By adopting the standardized \emph{OpenAI gym} interface, we hope to make it easy for researchers and practitioners to test their agents and algorithms on our challenges. In particular, we hope to simplify the process of deploying and training off-the-shelf RL agent, as well as provide interesting problems that may promote the development of new learning algorithms.

\section{Ethical considerations} \label{sec:Ethical}

RL agents trained for ethical penetration testing carry with them the potential for malicious misuse. In particular, the same agents may be deployed and adapted with the aim of generating material or immaterial damage. We would like to repeat that the aim of the current study is to develop agents to assist ethical hackers in legitimate penetration testing, and to develop an understanding of RL agents on a preventive ground only. For this reason, we advocate the development of agents in the context of CTF challenges, where the aim is a minimal and harmless exploitation of a vulnerability as a proof-of-concept (capture of the flag), but no further attacks are considered. We distance ourselves and condemn any application of these results for the development of offensive tools, especially in a military context\footnote{\url{https://futureoflife.org/open-letter-autonomous-weapons/}}.

\section{Conclusions} \label{sec:Conclusions}
In this paper we presented a model, named \emph{Agent Web Model}, that defines web hacking at different levels of abstraction. This formulation allows for a straightforward implementation of problems suited for machine learning agents. Since the aim and type of web attacks can be various, and different technical and human methods may be involved, we first restricted our attention to CTF-style hacking problems. 
We then modeled CTF-style web hacking as a game and as a RL problem. The RL problem considers a single player dealing with a static website consisting of objects with which the agent can interact by sending requests (with or without parameters). We formalized RL problems on 7 different levels of abstraction, ordered by increasing complexity in terms of number of objects, actions, parameters and states. Starting from a simple challenge on the first level of abstraction, we observed the complexity of the problems quickly increasing, thus defining a non-trivial learning challenge for an artificial agent.
An implementation of the problems on the first levels of abstraction was provided. The challenges we implemented range in complexity, they allow for customizability, and provide a way to instantiate a large number of random web hacking challenges in a generative way in order to train an artificial agent. Future work will be directed to further developing and further standardizing CTF challenges at higher levels of abstraction, as well as applying state of the art RL techniques to the problems we defined.

It is our hope that the formalization presented in this paper may not only allow for the development of automatic red bots that may help in the task of ethical penetration testing, but also promote the interaction and the research in both fields of machine learning and computer security: helping security expert to define realistic and relevant challenges that meet the formalism of machine learning, and offering to the RL expert stimulating problems that may foster advances in machine learning.

\bibliographystyle{spbasic}
\bibliography{ctf}

\end{document}